\documentclass[twocolumn]{aastex63}
\usepackage{CJK}
\usepackage{amsmath}

\received{}
\revised{}
\accepted{}
\submitjournal{ApJ}

\shorttitle{HCS and turbulence}
\shortauthors{Shi et al.}
\graphicspath{{./}{figures/}}

\begin{document}
\begin{CJK*}{UTF8}{gbsn}
\title{Influence of the heliospheric current sheet on the evolution of solar wind turbulence}

\correspondingauthor{Chen Shi}
\email{cshi1993@ucla.edu}

\author[0000-0002-2582-7085]{Chen Shi (时辰)}
\affiliation{Earth,Planetary, and Space Sciences, University of California, Los Angeles \\
    Los Angeles, CA 90095, USA}
    
\author[0000-0002-2381-3106]{Marco Velli}
\affiliation{Earth,Planetary, and Space Sciences, University of California, Los Angeles \\
    Los Angeles, CA 90095, USA}

\author[0000-0003-2880-6084]{Anna Tenerani}
\affiliation{Department of Physics, The University of Texas at Austin, \\
     TX 78712, USA}

\author[0000-0002-2916-3837]{Victor R\'eville}
\affiliation{IRAP, Universit\'e Toulouse III - Paul Sabatier,
CNRS, CNES, Toulouse, France}

\author[0000-0001-9030-0418]{Franco Rappazzo}
\affiliation{Earth,Planetary, and Space Sciences, University of California, Los Angeles \\
    Los Angeles, CA 90095, USA}



\begin{abstract}
The effects of the heliospheric current sheet (HCS) on the evolution of Alfv\'enic turbulence in the solar wind are studied using MHD simulations incorporating the expanding-box-model (EBM). The simulations show that near the HCS, the Alfv\'enicity of the turbulence decreases as manifested by lower normalized cross helicity and larger excess of magnetic energy.
The numerical results are supported by a superposed-epoch analysis using OMNI data, which shows that the normalized cross helicity decreases inside the plasma sheet surrounding HCS, and the excess of magnetic energy is significantly enhanced at the center of HCS. Our simulation results indicate that the decrease of Alfv\'enicity around the HCS is due to the weakening of radial magnetic field and the effects of the transverse gradient in the background magnetic field. The magnetic energy excess in the turbulence may be a result of the loss of Alfv\'enic correlation between velocity and magnetic field and the faster decay of transverse kinetic energy with respect to magnetic energy in a spherically expanding solar wind.
\end{abstract}



\section{Introduction} \label{sec:introduction}
Turbulence is a pervasive phenomenon in fluids and plasmas and has been observed to be a major feature of the solar wind. It is thought to be one of the main processes leading to solar wind heating \citep[e.g.][]{kiyani2015dissipation} and contribute to its acceleration from the solar corona \citep[e.g.][]{belcher1971alfvenic,leer1982acceleration}, while also affecting the acceleration and transport of energetic particles. Thus, understanding the origin and evolution of solar wind turbulence is crucial for fully understanding the heliosphere as a whole. 

\citet{Coleman1968}, using Mariner 2 data, showed that the power spectra of fluctuations in the solar wind have power-law scaling relations, indicating a well-developed turbulence. \citet{BelcherandDavis1971} found that these fluctuations are mainly outward-propagating Alfv\'en waves, with nearly-incompressible plasma density and magnetic field. An important question is how the Alfv\'enic fluctuation, or Alfv\'enic turbulence, evolves radially. Wentzel-Kramers-Brillouin (WKB) \citep[e.g.][]{AlazrakiandCouturier1971,belcher1971alfvenic,Hollweg1974} theory of the radial evolution of the Alfv\'en wave amplitude was developed. Non-WKB \citep[e.g.][]{HeinemannandOlbert1980,Barkhudarov1991,Velli1993} theory shows that linear coupling between the outward- and inward-propagating Alfv\'en waves leads to frequency-dependent reflection of the waves. Magnetohydrodynamic (MHD) models were developed on the evolution of the turbulence spectra and other parameters such as the wave energy densities \citep[e.g.][]{Tuetal1984,ZhouandMatthaeus1990a,Zanketal1996}. 

Els\"asser varibles $\mathbf{z}^\pm = \mathbf{u} \pm \mathbf{b}$, where $\mathbf{u}$ and $\mathbf{b}$ are velocity and magnetic field expressed in Alfv\'en speed, i.e. $\mathbf{b}= \hat{\mathbf{b}}/\sqrt{4\pi \rho}$ with $\hat{\mathbf{b}}$ and $\rho$ being the magnetic field and plasma density, are convenient in describing the Alfv\'enic fluctuations as they represent the inward and outward propagating Alfv\'en waves respectively. Two quantities have been used as important diagnostics of the Alfv\'enic turbulence, namely the normalized cross helicity ($\sigma_c$) defined as: 
\begin{equation}\label{eq:sigma_c}
    \sigma_c = \frac{\left| \mathbf{z^-} \right|^2 - \left| \mathbf{z^+} \right|^2}{\left| \mathbf{z^-} \right|^2 + \left| \mathbf{z^+} \right|^2},
\end{equation}
and normalized residual energy ($\sigma_r$) defined as:
\begin{equation}\label{eq:sigma_r}
    \sigma_r = \frac{\left| \mathbf{u} \right|^2 - \left| \mathbf{b} \right|^2}{\left| \mathbf{u} \right|^2 + \left| \mathbf{b} \right|^2}.
\end{equation}
$\sigma_c$ measures the relative amount of outward and inward wave energies and $\sigma_r$ measures the relative amount of kinetic and magnetic energies. A large number of works were conducted on the radial evolution of these two quantities in the solar wind \citep[e.g.][]{Robertsetal1987,Bavassanoetal1998,bruno2007magnetically,Chenetal2020,shi2021alfvenic} but two outstanding problems remain unresolved. First, the normalized cross helicity decreases with radial distance to the Sun, and second, a prevailing negative value of residual energy is observed. It is known that, in a homogeneous medium with a uniform background magnetic field, the Alfv\'enic turbulence evolves toward a status that only one wave population survives, i.e. $\sigma_c = \pm 1$. This is the so called ``dynamic alignment'' \citep{Dobrowolny1980}. In contrast to the theory, in the solar wind, the dominance of the outward Alfv\'en wave is gradually weakened during the radial propagation as manifested by the decrease of $|\sigma_c|$. Besides, in a purely Alfv\'enic system, the kinetic and magnetic energies should be exactly equi-partitioned, i.e. $\sigma_r=0$, while in the solar wind, a magnetic energy excess is typically observed even at distances very close to the Sun, below 30 solar radii \citep{Chenetal2020,mcmanus2020cross,shi2021alfvenic}.

\begin{figure*}[htb!]
    \centering
    \includegraphics[width=\textwidth]{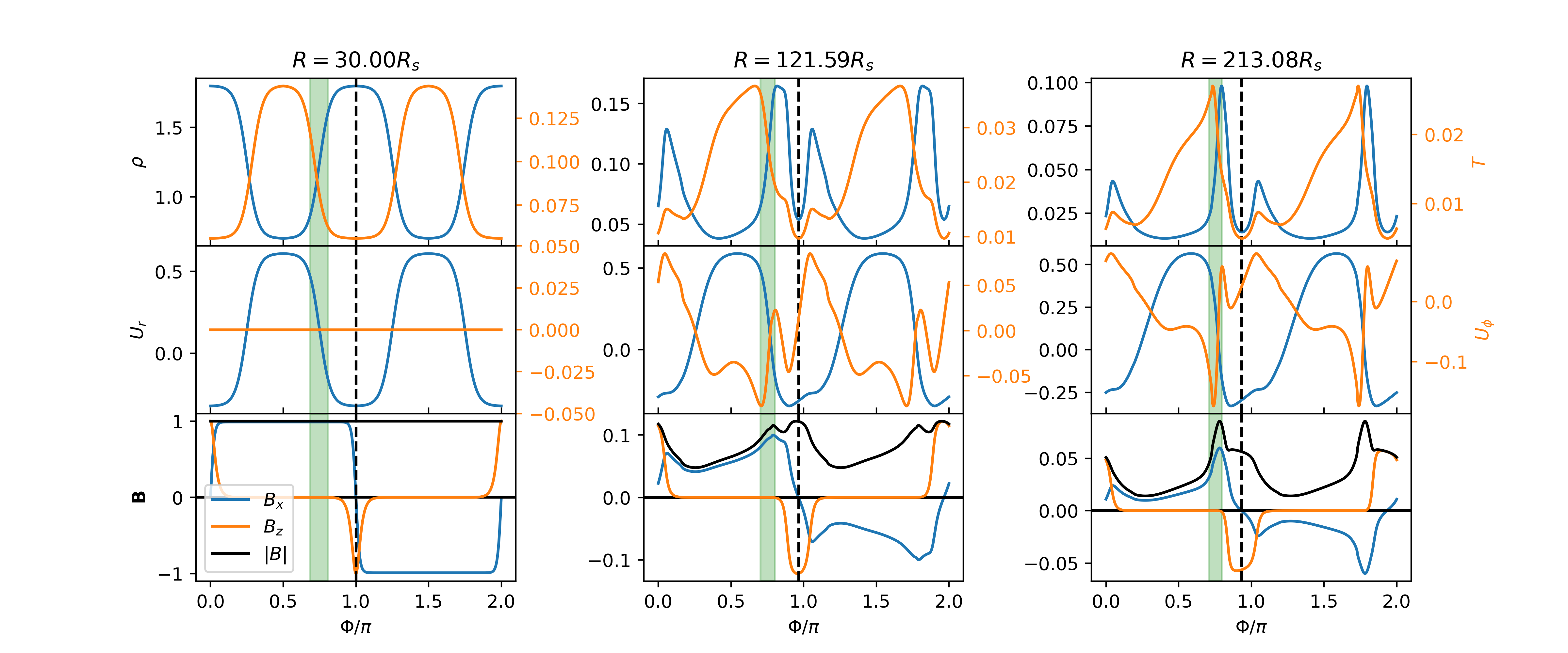}
    \caption{Evolution of the longitudinal profiles of background fields from a 1D run without any waves. Left column is the initial status at $30R_s$, middle column is at 121.59$R_s$, and right column is around 1 AU. Top row: density (blue) and temperature (orange). Middle row: radial speed (blue) in the expanding-box frame and longitudinal speed (orange). Bottom row: radial component (blue), out-of-plane component (orange), and magnitude (black) of the magnetic field. The quantities are normalized (see the text). The green shade shows the compression region and the vertical dashed line marks the polarity reversal of the magnetic field.}
    \label{fig:evolution_background_field}
\end{figure*} 

One possible mechanism that resolves these paradoxes is the large-scale velocity shear in the solar wind. \citet{Coleman1968} proposed that the differential streaming generates Alfv\'en waves at long wavelengths. These newly-generated waves do not have a preferential propagating direction and thus the initial dominance of the outward wave gradually declines. This shear-driven decrease of cross helicity was confirmed by numerical simulations \citep{Robertsetal1992,Shietal2020}. Velocity shear is widely adopted in turbulence models as a source for the wave energies and is able to reproduce the observed decrease of $\sigma_c$ in the models. On the contrary, there has been no satisfactory solar wind turbulence model that leads to negative residual energy so far. For example, in the model by \citep{Zanketal2017}, the source term for the residual energy is attributed to the stream shear but whether this term causes growth or decay of the residual energy is arbitrary. Heliospheric current sheet (HCS) is a good candidate that may influence the evolution of Alfv\'enic turbulence. It is shown by both simulations and in-situ observations that in the proximity of the HCS, the Alfv\'enicity of the turbulence decreases in general \citep[e.g.][]{goldstein1999magnetohydrodynamic,chen2021near}. However, how the HCS modifies the dynamic evolution of $\sigma_c$ and $\sigma_r$ in the spherically expanding solar wind is still unclear. 

In this study, we carry out two-dimensional MHD simulations using the expanding-box-model (EBM). Large-scale solar wind structures, including the fast-slow stream interaction regions (SIRs) and the HCS, are constructed and evolve self-consistently in the simulations. We investigate how properties of the Alfv\'enic turbulence evolve radially and how SIRs and HCS modify its evolution. A superposed-epoch analysis of HCS crossings at 1 AU is carried out using the OMNI dataset and the turbulence properties near the HCS are examined. The paper is organized as follows. In Section \ref{sec:EBM_sim} we present the setup of the MHD simulations and the numerical results. In Section \ref{sec:sup_ep_ana} we present the superposed-epoch analysis of the HCS crossings observed at 1 AU. We then discuss our results in Section \ref{sec:discussion} and conclude in Section \ref{sec:conclusion}.


\begin{figure*}[ht!]
    \centering
    \includegraphics[scale=0.7]{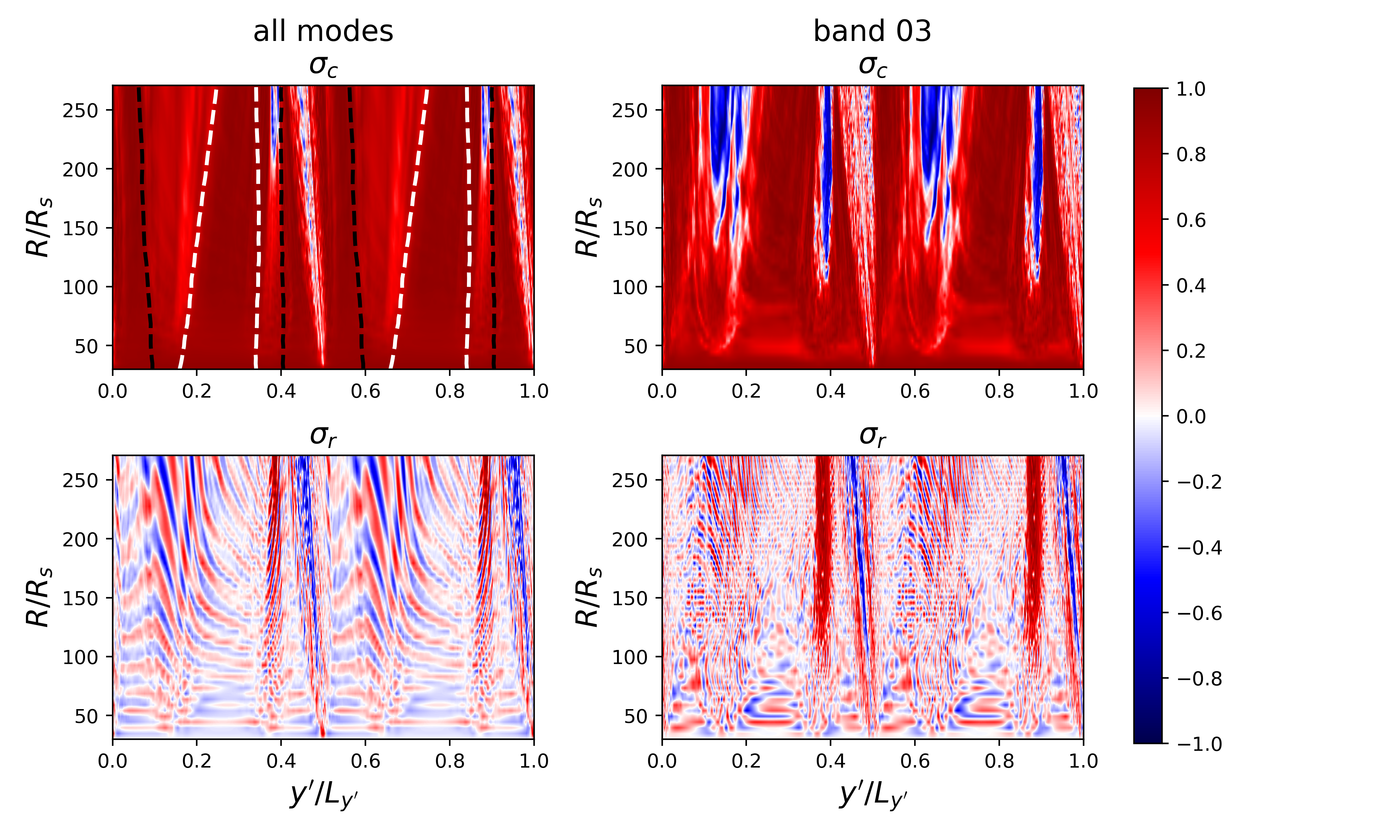}
    \caption{$y^\prime-R(t)$ profiles color-coded with the normalized cross helicity $\sigma_c$ (top row) and the normalized residual energy $\sigma_r$  (bottom row). Left column is the result using wave energies integrated over all modes and right column is the result for wave modes $[4,8)$, or wave length $\lambda \in [2.5R_s, 1.25R_s)$, corresponding roughly to wave period $T\in [38,75)$min. In the top-left panel, the dashed white lines mark the boundaries of fast streams, defined as $y^\prime(t)$ at which the background radial speed equals 650 km/s, and the dashed black lines mark the boundaries of slow streams, defined as $y^\prime(t)$ at which the background radial speed equals 400 km/s.}
    \label{fig:sigma_c_sigma_r_sim}
\end{figure*}

\section{Expanding-Box-Model simulation}\label{sec:EBM_sim}
\subsection{Numerical method and 1D test run}\label{sec:numerical_method}
The code we use for simulations is a 2D pseudo-spectral MHD code used by \citet{Shietal2020} to study the interaction between SIRs and turbulence. EBM module is implemented so that the radial expansion effect of the solar wind is taken into account \citep{Grappinetal1993,GrappinandVelli1996}. The expansion effect is not negligible in the solar wind because it induces inhomogeneity of the background streams, which leads to the reflection of the Alfv\'en waves \citep[e.g.][]{HeinemannandOlbert1980}, and anisotropic evolution of velocity and magnetic fields in the radial and transverse directions \citep[e.g.][]{dong2014evolution}. The EBM has also been employed to reproduce the ``magnetic switchbacks'' observed in the young solar wind \citep{squire2020situ}. In our code, a finite spiral angle of magnetic field and stream interface is allowed so that compression and rarefaction regions are constructed. A detailed description of the numerical method can be found in \citet{Shietal2020}.

We carried out four simulations with two free parameters, namely whether a non-zero spiral angle is set and whether the expansion effect is included. Here we mainly present results from the run with both a non-zero spiral angle and the expansion effect, i.e. the most realistic run. The parameters for the initial setup are chosen according to solar wind observations as described below. The simulation starts from $R_0 = 30R_s$ and ends at $R = 270.9 R_s$ where $R_s$ is the solar radius. The initial spiral angle is $\alpha = 8.1^\circ$ so that the angle becomes $45^\circ$ at 1 AU. The size of the simulation domain is $L_{x^\prime} \times L_{y^\prime} = 10R_s \times 2\pi R_0$ where $x^\prime - y^\prime$ is the corotating coordinate system, i.e. $x^\prime$ is parallel to the background magnetic field and $y^\prime$ is the quasi-longitudinal direction \citep{Shietal2020}. $L_{y^\prime} = 2 \pi R_0$ means that the domain is a full circle in the ecliptic plane. The initial background fields as functions of the quasi-longitudinal direction $y^\prime$ are plotted in the left column of Figure \ref{fig:evolution_background_field} with normalized units. The quantities for normalization are: $\bar{n}=200$ cm$^{-3}$, $\bar{B}=250$ nT, $L=R_s$, $\bar{V}=\bar{B}/\sqrt{4\pi m_p \bar{n}}=386$ km/s, and $\bar{P}=\bar{B}^2/4\pi = 49.7$ nT. The wind consists of two fast streams and two slow streams. The transition between fast and slow streams is of the form $\tanh{\left(y^\prime/a_{sh}\right)}$ where $a_{sh} = 0.075 \pi R_0$. The velocity of the streams is purely radial with the speed ranging from 340 km/s to 700 km/s. We note that in Figure \ref{fig:evolution_background_field} the velocity is in the reference frame of the expanding box, which moves radially with the average speed of the plasma inside the simulation domain $U_0=464$ km/s. The density is 360 $\mathrm{cm}^{-3}$ for the slow streams and 140 $\mathrm{cm}^{-3}$ for the fast streams. Two Harris current sheets with thickness $a_{cs}=0.025 \pi R_0$ are embedded in the center of the slow streams. The current sheets are force-free, as the in-situ measurement of the HCS usually shows \citep{Smith2001}, and a finite $z$-component is used to make sure that $\left|\mathbf{B} \right|$ is constant and the magnetic field strength is uniformly $250$ nT. The pressure is uniform and equals 5 nPa, i.e. the plasma beta is $\beta=8\pi p/B^2 \approx 0.2$, so that the temperatures of the fast and slow streams are $2.6$ MK and $1.0$ MK respectively. The adiabatic index is set to be $3/2$ instead of $5/3$ since the solar wind observations show that the cooling rate of the solar wind is slower than a $5/3$ adiabatic cooling \citep[e.g.][]{Hellingeretal2011}.

\begin{figure*}[ht!]
    \centering
    \includegraphics[scale=0.75]{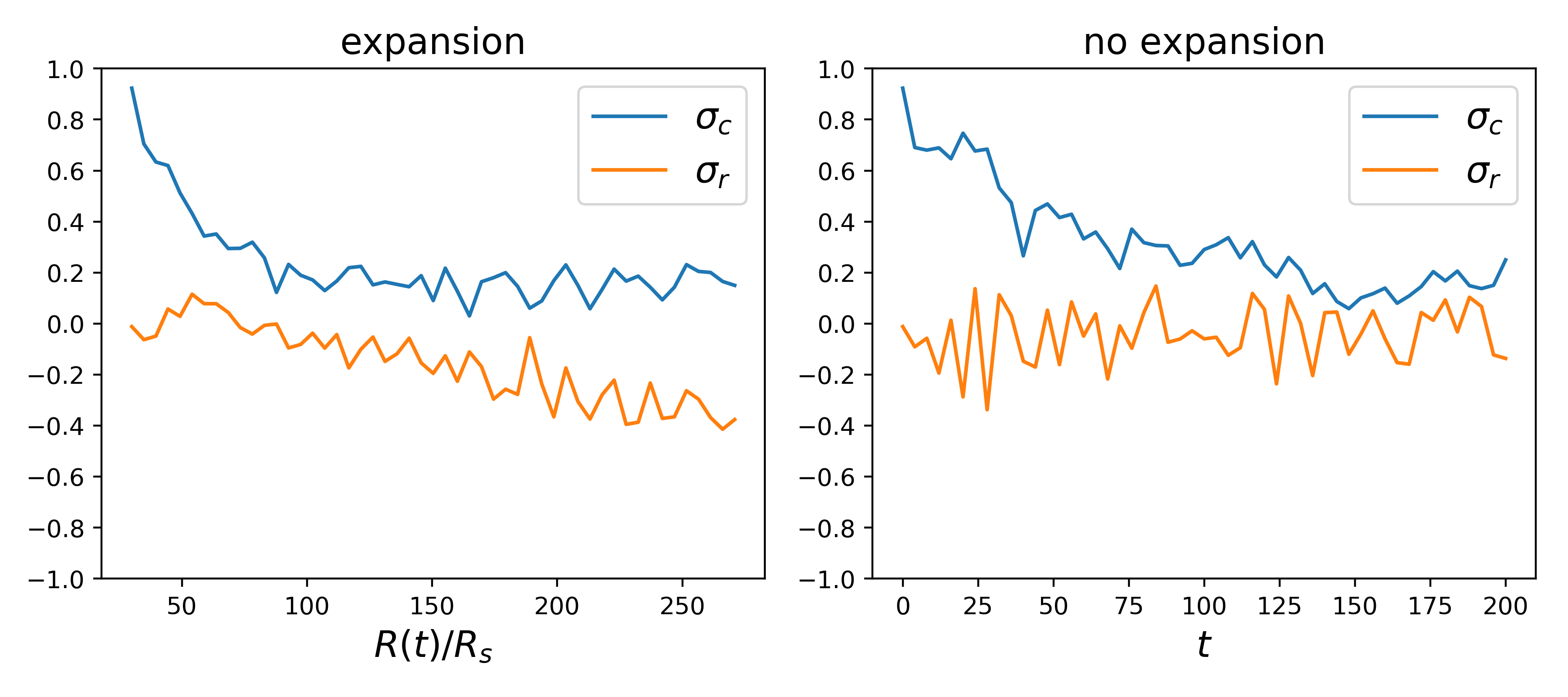}
    \caption{Left: Radial evolution of $\sigma_c$ (blue) and $\sigma_r$ (orange) in a longitude band with thickness $2a_{cs}$ around the current sheet in the run with expansion (Figure \ref{fig:sigma_c_sigma_r_sim}). Right: Similar to left panel, time evolution of $\sigma_c$ (blue) and $\sigma_r$ (orange) near the current sheet in the run without expansion.}
    \label{fig:sigma_c_sigma_r_radius_CS}
\end{figure*}

In Figure \ref{fig:evolution_background_field}, we show the evolution of the background fields from a 1D run with grid points on $y^\prime$-axis, i.e. the quasi-longitude direction. Top row shows density (blue) and temperature (orange), middle row shows the radial component (blue) and transverse component (orange) of the velocity, bottom panel shows the radial component (blue), out-of-plane component (orange), and the magnitude (black) of the magnetic field. From left to right columns are the profiles of the background fields at three locations as labeled on the top. In each column, the green shade shows one of the two compression regions and the vertical dashed line marks one of the polarity reversal points of the radial magnetic field. We can see that, as the simulation proceeds, a compression region with enhanced density, temperature, and magnetic field strength forms between the trailing fast stream and leading slow stream. A divergent transverse flow due to the deflection of the streams is also observed in the compression region. In the rarefaction region, i.e. the trailing edge of the fast slow, the density, temperature, and magnetic field strength decrease. But we note that near the current sheet, a dip is seen in density and temperature but the magnetic field strength shows a peak. This is a result of the differential decay rates of $B_z$ and $B_x$ induced by the expansion of the solar wind. As $B_z \propto 1/R$ decays slower than $B_x \propto 1/R^2$, an initial force-free current sheet cannot maintain a uniform magnetic pressure once the expansion starts and a local peak of magnetic pressure forms at the polarity reversal point. The large magnetic pressure will then push the plasma away from the current sheet, resulting in a local dip in density and temperature and also a divergent transverse flow.


\begin{figure*}[ht!]
    \centering
    \includegraphics[width=\textwidth]{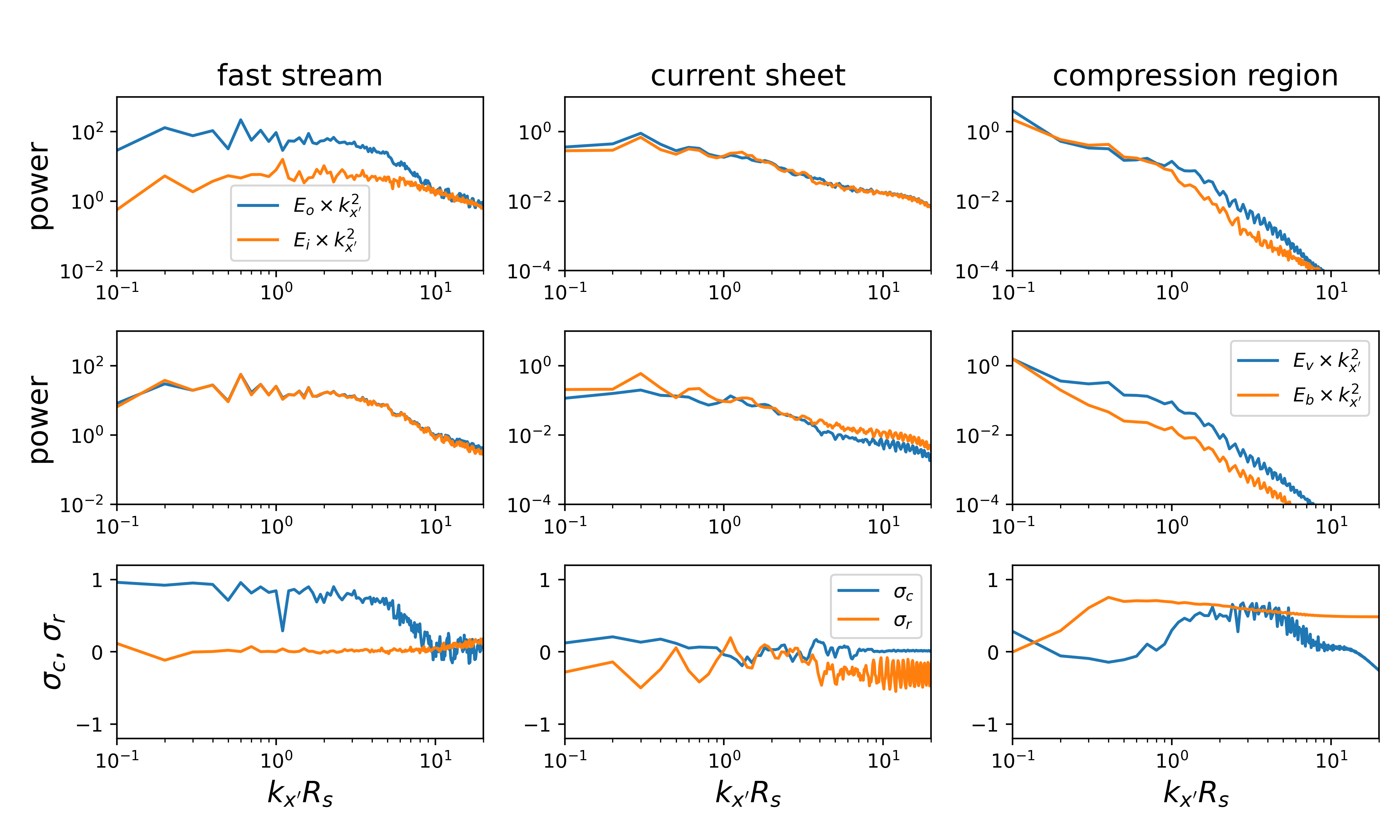}
    \caption{$x^\prime$-spectra of various fields at $R(t)=217.9R_s$ inside the fast stream (left column), the current sheet (middle column) and the compression region (right column). The spectra are averaged in a $y^\prime$-band of width $2a_{cs}$ inside each region. Top row: spectra of the outward (blue) and inward (orange) Alfv\'en waves (Els\"asser variables). Middle row: spectra of the kinetic (blue) and magnetic (orange) perturbations. Bottom row: spectra of $\sigma_c$ (blue) and $\sigma_r$ (orange) calculated from the spectra in top and middle rows.}
    \label{fig:spectra_sim}
\end{figure*}

\subsection{Results of the 2D runs with waves}
Based on the 1D run, 2D simulations are conducted. At initialization, circularly-polarized Alfv\'enic wave bands comprising 16 modes are added on top of the background streams with wave vectors parallel to $\hat{e}_{x^\prime}$ and correlated $\mathbf{u}$ \& $\mathbf{b}$ fluctuations in $y^\prime$ \& $z$ directions (see \citep{Shietal2020} for more details). The energy spectrum of the initial waves obeys $E(k_{x^\prime}) \propto k_{x^\prime}^{-1}$ with $k_{x^\prime,m} = m/L_{x^\prime}$ where $m=1,2,\cdots 16$, i.e. the longest and shortest wavelengths are $10R_s$ (wave period $T \approx \lambda/U_0=4$ hr) and $0.625R_s$ (wave period $T \approx \lambda/U_0=15$ min) respectively. Both outward and inward waves are added with the amplitude of inward waves being $1/5$ of the amplitude of outward waves. The amplitude of the first mode of outward waves is $0.2B_0=50$ nT. To avoid sharp jumps of perturbation fields across the current sheets, we modulate the wave amplitudes by a hyperbolic tangent function, similar to the $x^\prime$-component of the background magnetic field, such that the wave amplitudes are exactly 0 at the center of the current sheets. Note that this requires non-zero wave components along $x^\prime$ inside the current sheets to ensure $\nabla \cdot \mathbf{B} = 0$. The number of grid points is $n_{x^\prime} \times n_{y^\prime} = 2048 \times 8192$ (we note that the simulation domain size is $10R_s\times 60 \pi R_s$) so that the smallest wavelength that is resolved is $\lambda = 2 \Delta x^\prime \approx 0.01 R_s$, corresponding to a wave period $T \approx \lambda/U_0 \approx 15$s, which is approximately two magnitudes larger than the ion gyro-scale and ion inertial scale. But we note that in the MHD simulations, there are no intrinsic gyro-scales and inertial scales, or in other words these kinetic scale lengths are zero in MHD.

We process the simulation data using the same method as described in \citep{Shietal2020}. At each time, or radial distance to the Sun equivalently, we first calculate the $x^\prime$-averaged fields, i.e. the background fields. Then we remove the background fields to get the wave fields. We only analyze wave components that are perpendicular to the $x-y$ plane background magnetic field $\mathbf{B_0}=B_0\left( y^\prime \right) \hat{e}_{x^\prime}$. We note here that in the $x-y$ plane, the background magnetic field is always aligned with $\hat{e}_{x^\prime}$ as the $x^\prime$ axis rotates away from the radial direction as the solar wind expands \citep{Shietal2020}. The perturbed Els\"asser variables are defined by
\begin{equation}
    \mathbf{z_{out}} = \mathbf{u_1} - \mathrm{sign}(B_0) \frac{\mathbf{b_1}}{\sqrt{\rho}} , \, \mathbf{z_{in}} = \mathbf{u_1} + \mathrm{sign}(B_0) \frac{\mathbf{b_1}}{\sqrt{\rho}}.
\end{equation}
We then apply Fourier transform along $x^\prime$ to $\mathbf{u}_1$, $\mathbf{b}_1/\sqrt{\rho}$, $\mathbf{z_{out}}$ and $\mathbf{z_{in}}$. As there are 2048 grid points along $x^\prime$, 1024 wave modes are resolved with wave-numbers $k_{x^\prime}=(1,2,\cdots 1024)/L_{x^\prime}$, i.e. mode $m$ corresponds to $k_{x^\prime}=m/L_{x^\prime}$. We divide all the wave modes into 10 wave-number bands which are logarithmically spaced, i.e. band $i$ contains modes $\left[2^{i-1}, 2^i\right)$, or wavelengths between $L_{x^\prime}/2^{i-1}$ and $L_{x^\prime}/2^{i}$. We then calculate $\sigma_c(y^\prime)$ and $\sigma_r(y^\prime)$ for each wave-number band and also for integration of all wave modes. 

In Figure \ref{fig:sigma_c_sigma_r_sim} we present the $y^\prime-R(t)$ profiles color-coded with $\sigma_c$ (top row) and $\sigma_r$ (bottom row) for all wave modes (left column) and for band 03 (right column) which corresponds to wave length $\lambda \in \left[2.5R_s, 1.25R_s \right)$, or roughly wave period $T\in [38,75)$min. The figure is produced by piling up the $y^\prime$ profiles of $\sigma_{c,r}$ at different moments or equivalently different radial locations $R(t)$. In top-left panel, we mark the boundaries of fast streams, defined as $y^\prime(t)$ at which the background radial speed equals 650 km/s, by dashed white lines and mark the boundaries of slow streams, defined as $y^\prime(t)$ at which the background radial speed equals 400 km/s by dashed black lines. The two current sheets are located in the center of the slow streams, around $y^\prime = 0.5 L_{y^\prime}$ and $y^\prime = L_{y^\prime}$. We first inspect the left column of Figure \ref{fig:sigma_c_sigma_r_sim} which shows $\sigma_c$ and $\sigma_r$ calculated using wave energies integrated over all wave-numbers. As already discussed by \citet{Shietal2020}, in regions with nearly uniform background fields, i.e. inside fast streams and inside slow streams far from the current sheets, $\sigma_c$ remains almost constant throughout the evolution, indicating that the Alfv\'enicity remains high in these regions. In the velocity-shear regions (regions between the black and white lines), $\sigma_c$ declines with radial distance. Especially, in the compression region (around $y^\prime =(0.35-0.4) L_{y^\prime}$ and $y^\prime =(0.85-0.9) L_{y^\prime}$), $\sigma_c$ drops below 0 beyond $200R_s$. \citet{Shietal2020} showed that in shear regions, the damping of the outward Alfv\'en wave is significantly faster than the inward Alfv\'en wave, leading to the decrease in $\sigma_c$. Except for near the current sheets, which will be discussed in detail later, $\sigma_r$ oscillates around 0 in all regions, indicating that the Alfv\'enicity of the waves is well conserved for the long wavelength modes (we note that the integrated wave energies are dominated by the modes of largest scales). The oscillation in $\sigma_r$ is caused by periodic correlation and de-correlation between the outward and inward waves. From the left column of Figure \ref{fig:sigma_c_sigma_r_sim}, we also see that the evolution of Alfv\'en waves is significantly modified by the current sheets. In the neighborhood of the current sheets, $\sigma_c$ decreases quite fast and $\sigma_r$ evolves toward negative values. In the left panel of Figure \ref{fig:sigma_c_sigma_r_radius_CS} we plot the radial evolution of $\sigma_c$ and $\sigma_r$ in a band of width $2a_{cs}$ around the current sheet initially located at $y^\prime = 0.5 L_{y^\prime}$. $\sigma_c$ starts from a high value, i.e. 0.92 determined by the initial condition, drops to around 0.2 within 100$R_s$ and then remains stable. $\sigma_r$ starts from exactly 0, rises slightly at the beginning due to the increase of kinetic energy caused by the magnetic pressure gradient at the current sheet as discussed in Section \ref{sec:numerical_method}, and then starts to drop continuously, reaching a value $-0.3$ at 1 AU. For comparison, we plot the time evolution of $\sigma_c$ and $\sigma_r$ from the run without expansion in the right panel of Figure \ref{fig:sigma_c_sigma_r_radius_CS}. Evolution of $\sigma_c$ does not show much difference between the two runs while $\sigma_r$ remains around 0 in the run without expansion, indicating that expansion effect is important to the decrease of $\sigma_r$ around the current sheet, which will be discussed in more detail in Section \ref{sec:discussion}.

In the right column of Figure \ref{fig:sigma_c_sigma_r_sim}, we show the $y^\prime-R(t)$ profiles of $\sigma_c$ and $\sigma_r$ for wave band 03. Compared with the left column, the drop of $\sigma_c$ in velocity-shear regions is much more significant and the $\sigma_c$-drop regions around the current sheets are wider. For $\sigma_r$, the most prominent feature is that inside the compression regions, $\sigma_r$ evolves toward $+1$, i.e. kinetic energy becomes dominant in these regions, indicating that the large-scale velocity shear and compression facilitate the transfer of kinetic energy toward small scales.

In Figure \ref{fig:spectra_sim} we show the $x^\prime$-spectra, i.e. the parallel spectra, of various fields calculated at the moment $R(t)=217.9R_s$. From left to right columns are spectra averaged in $y^\prime$-bands of width $2a_{cs}$ inside fast stream, current sheet and compression region respectively. Top row shows the spectra of outward (blue) and inward (orange) Els\"asser variables. Middle row shows the spectra of kinetic (blue) and magnetic (orange) perturbations. We multiply these spectra by $k_{x^\prime}^{2}$ as the ``critical balance'' model \citep{goldreich1995toward} predicts a parallel spectrum $E \propto k_{\parallel}^{-2}$. We can see from Figure \ref{fig:spectra_sim} that in general these spectra are steeper than the prediction of the critical balance model except for inside the fast stream. This is because the shears in magnetic field and velocity turn the wave vector from quasi-parallel to quasi-perpendicular and enlarges the perpendicular wavenumber gradually which speeds up the dissipation of the wave energies \citep{Shietal2020}. Bottom row shows the spectra of $\sigma_c$ and $\sigma_r$ calculated from the spectra shown in top and middle rows. Inside the fast stream, $\sigma_c$ is close to 1 and $\sigma_r$ is close to 0 for most modes, except for close to the numerical dissipation range, meaning that the waves maintain a high Alfv\'enicity over a large span of wave numbers. Around the current sheet, $\sigma_c$ decreases to nearly 0 for all wave numbers. $\sigma_r$ is negative for most of the wave numbers, except for an intermediate range ($1<k_{x^\prime}R_s<3$) where it is around 0. In the compression region, $\sigma_c$ is overall smaller than the initial condition 0.92 but the curve of $\sigma_c$ shows a decrease with $k_{x^\prime}$ at small wave numbers and rises again. $\sigma_r$ is around 0 for small wave numbers and shows a significant increase with $k_{x^\prime}$, reaching its maximum value at the same $k_{x^\prime}$ where $\sigma_c$ reaches its local minimum. It indicates that in the compression region the large scale stream structure generates small-scale fluctuations that are kinetic energy dominated, as observed in previous simulations \citep[e.g.][]{Robertsetal1992}. The newly-generated fluctuations weaken the dominance of the outward Alfv\'en waves, consistent with the scenario proposed by \citet{Coleman1968}. We note that for large wave-numbers ($k_{x^\prime}R_s \gtrsim 5$) the spectra are significantly modulated by the explicit numerical filter applied to the simulation, thus spectral breaks can be seen at large wave-numbers.

We do not present the perpendicular power spectra for the following reasons. First, as the simulation coordinate system is non-Cartesian, there is no axis perpendicular to $\mathbf{B_0}$. The $y^\prime$ axis is perpendicular to $\mathbf{B_0}$ initially but the angle between $y^\prime$ and $\mathbf{B_0}$ gradually increases \citep{Shietal2020}. Second, because of the elongated simulation domain along $y^\prime$ and also due to the spherical expansion, the resolution in $y^\prime$ is much lower than that in $x^\prime$. Besides, as we would like to focus on certain longitudinal regions, e.g. around the HCS, instead of the whole $y^\prime$ range, the number of data points is limited. Because of the above reasons, it is difficult to produce physically meaningful perpendicular spectra.

\begin{figure}
    \centering
    \includegraphics[width=\hsize]{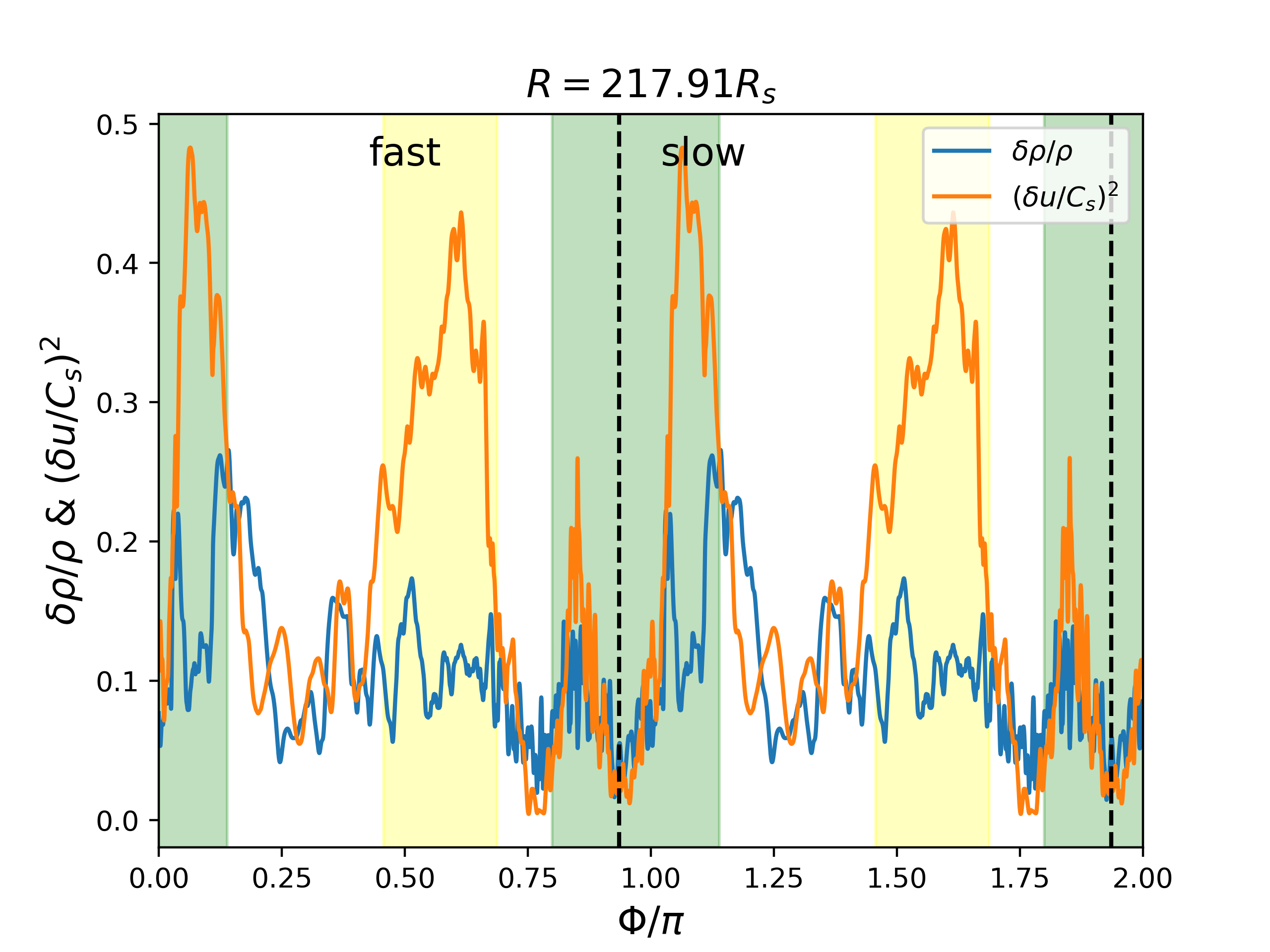}
    \caption{Longitude-profiles of the normalized density fluctuation $\delta \rho/\rho$ (blue) and the square of Mach number $(\delta u / C_s)^2$ (orange) in the simulation at moment $R(t)=217.91R_s$. $\delta \rho$ and $\delta u$ are the root-mean-squares of density and velocity calculated long $x^\prime$. $C_s$ is the sound speed. Yellow and green shades mark the fast and slow streams respectively. The vertical dashed lines are the locations of the current sheets.}
    \label{fig:delta_rho_mach_number_sim}
\end{figure}

In Figure \ref{fig:delta_rho_mach_number_sim}, we plot the $y^\prime$-profiles of the normalized density fluctuation $\delta \rho/\rho$ (blue) and the square of Mach number $(\delta u / C_s)^2$ (orange) in the simulation at moment $R(t)=217.91R_s$. Here $\delta \rho$ and $\delta u$ are the root-mean-squares of density and velocity calculated long $x^\prime$. $C_s$ is the sound speed. Yellow and green shades mark the fast and slow streams respectively. The vertical dashed lines are the locations of the current sheets. The normalized density fluctuation is overall small, mostly below 0.2, similar to the solar wind observation \citep[e.g.][]{shi2021alfvenic}. In the compression region and near the current sheet, e.g. from $\Phi = 0.7-1.1$, $\delta \rho /\rho$ is highly correlated with $(\delta u / C_s)^2$, implying a large compressive component in the velocity fluctuations in these regions.


\section{Superposed-epoch analysis of HCS at 1 AU}\label{sec:sup_ep_ana}
Although works have been carried out on turbulence properties around SIRs \citep[e.g.][]{BorovskyandDenton2010}, literature on how HCS affects the solar wind turbulence is still incomplete. To validate our simulation results, we carry out a superposed-epoch analysis of HCS crossings at 1 AU and study how the properties of turbulence change near HCS.

\subsection{Selection and structure of HCS}
\begin{figure}[ht!]
    \centering
    \includegraphics[scale=0.4]{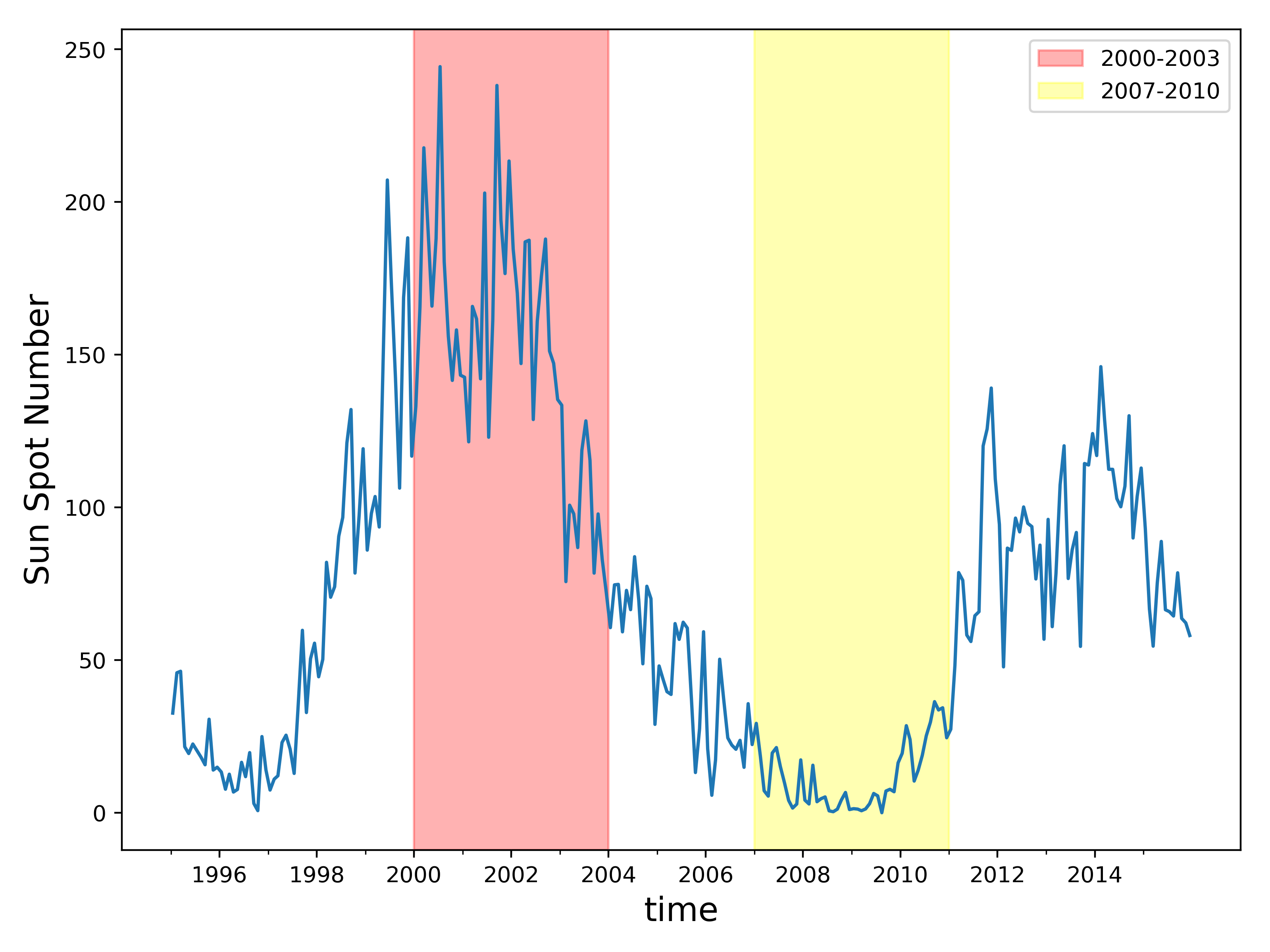}
    \caption{Monthly sunspot number from 1995 to 2015. The two shaded regions mark the two time periods used for OMNI data analysis and correspond to solar maximum (2000-2003) and solar minimum (2007-2010) respectively.}
    \label{fig:sunspot_number}
\end{figure}

For the current study, we use the OMNI dataset, which contains magnetic field and plasma data from multiple spacecraft, including ACE and WIND \citep{KingandPapitashvili2005}. Time resolution of the data is 1 minute, enough for study of MHD turbulence. We analyze data during two 4-year periods: 2000-2003 which is around solar maximum of solar cycle 23, and 2007-2010 which is around solar minimum between solar cycles 23 \& 24, as shown by the shaded regions in Figure \ref{fig:sunspot_number}, which plots monthly sunspot number.

\begin{figure*}[htb!]
    \centering
    \includegraphics[width=\textwidth]{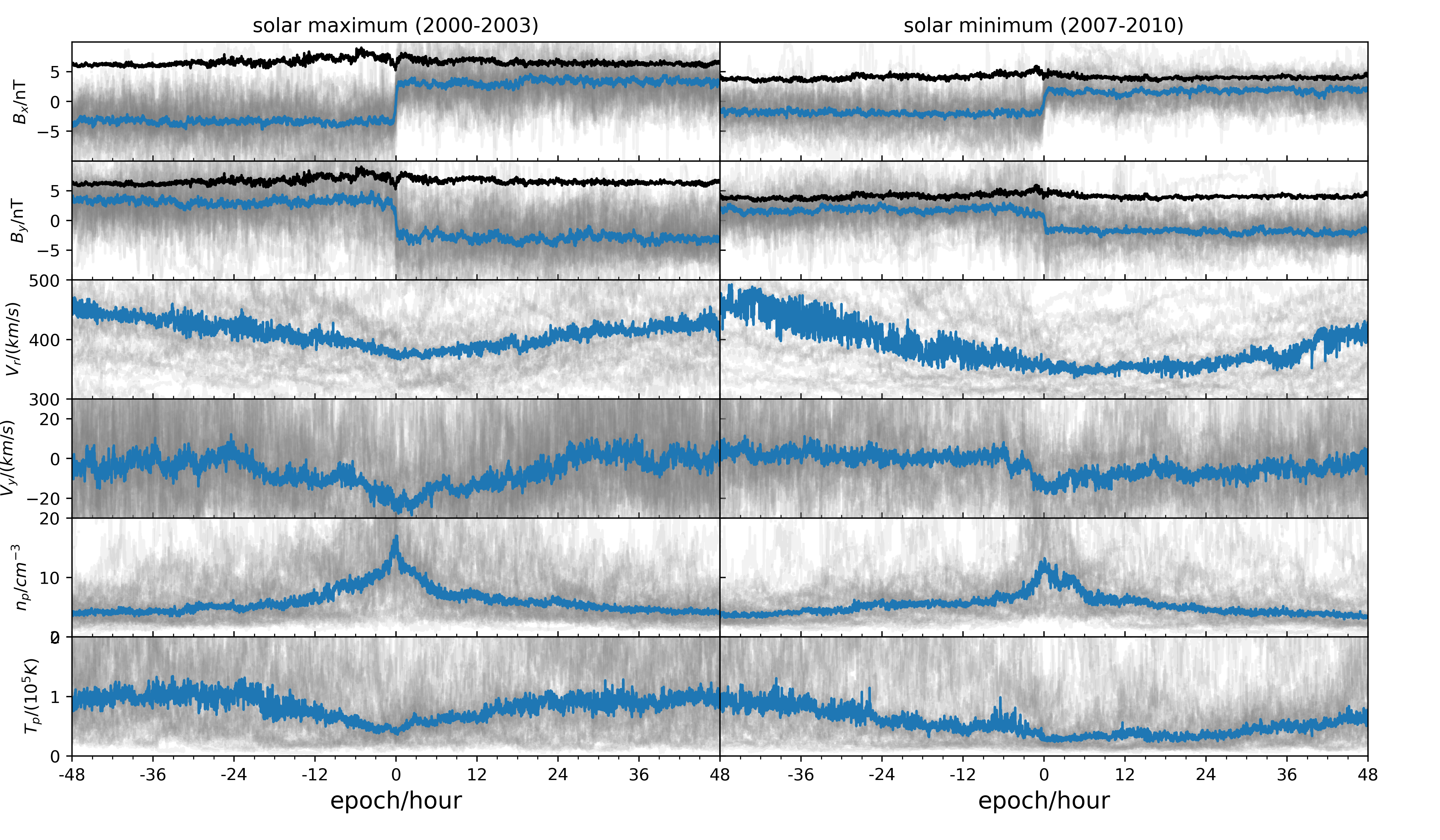}
    \caption{Superposed-epoch analysis of the HCS crossings. Epoch 0 is the moment of crossing. From top to bottom rows are $B_x$ and $B_y$ in GSE coordinates, $V_r$, GSE $V_y$, proton density and proton temperature respectively. Left column is solar maximum and right column is solar minimum. In each panel, grey lines are individual events and blue line is the median value. In the top two rows, the black curves are the medians of $|\mathbf{B}|$.}
    \label{fig:HCS_structure}
\end{figure*}

The procedure to select HCS crossings is stated as follows. We first calculate the one-day average of $B_{x,GSE}$, which is equivalently the opposite of radial component of the solar wind magnetic field. Then we find days when its polarity changes and we require that the polarities before and after each polarity-reversal day maintain at least 4 days. Then we inspect the 1-minute data to determine the exact polarity reversal times. We identify 48 events for solar maximum and 45 events for solar minimum. List of the HCS crossings is shown in Table \ref{tab:HCS_crossing}.

Superposed-epoch analysis of the HCS structure is shown in Figure \ref{fig:HCS_structure}. From top to bottom rows are GSE $B_x$, GSE $B_y$, $V_r$, GSE $V_{y}$, proton density, and proton temperature respectively. Left column is solar maximum and right column is solar minimum. In each panel, grey curves are individual events and blue curve is the median value of all events. In the top two rows we also plot medians of the magnetic field strength in black curves. We have reversed the time series of certain events such that $B_x$ is always changing from negative to positive. 
The time scale for HCS crossings is on average 1-2 hours and the HCS is embedded in much thicker (1-2 days) plasma sheets with enhanced proton density and lower proton temperature. The magnetic field strength is quite constant across the HCS, implying a force-free structure. By comparing left and right columns of Figure \ref{fig:HCS_structure}, we see that the strength of magnetic field is larger in solar maximum than solar minimum. Another thing to notice is that there is a negative GSE $V_y$ at the HCS, i.e. the plasma flow is rotating in the same direction with the solar rotation. The reason might be that HCS is usually embedded in slow solar wind ahead of the compression region and are pushed along longitudinal direction in accordance with solar rotation \citep{EselevichandFilippov1988,Siscoe1972}.

\subsection{Turbulence properties near HCS}

\begin{figure*}[ht!]
    \centering
    \includegraphics[scale=0.5]{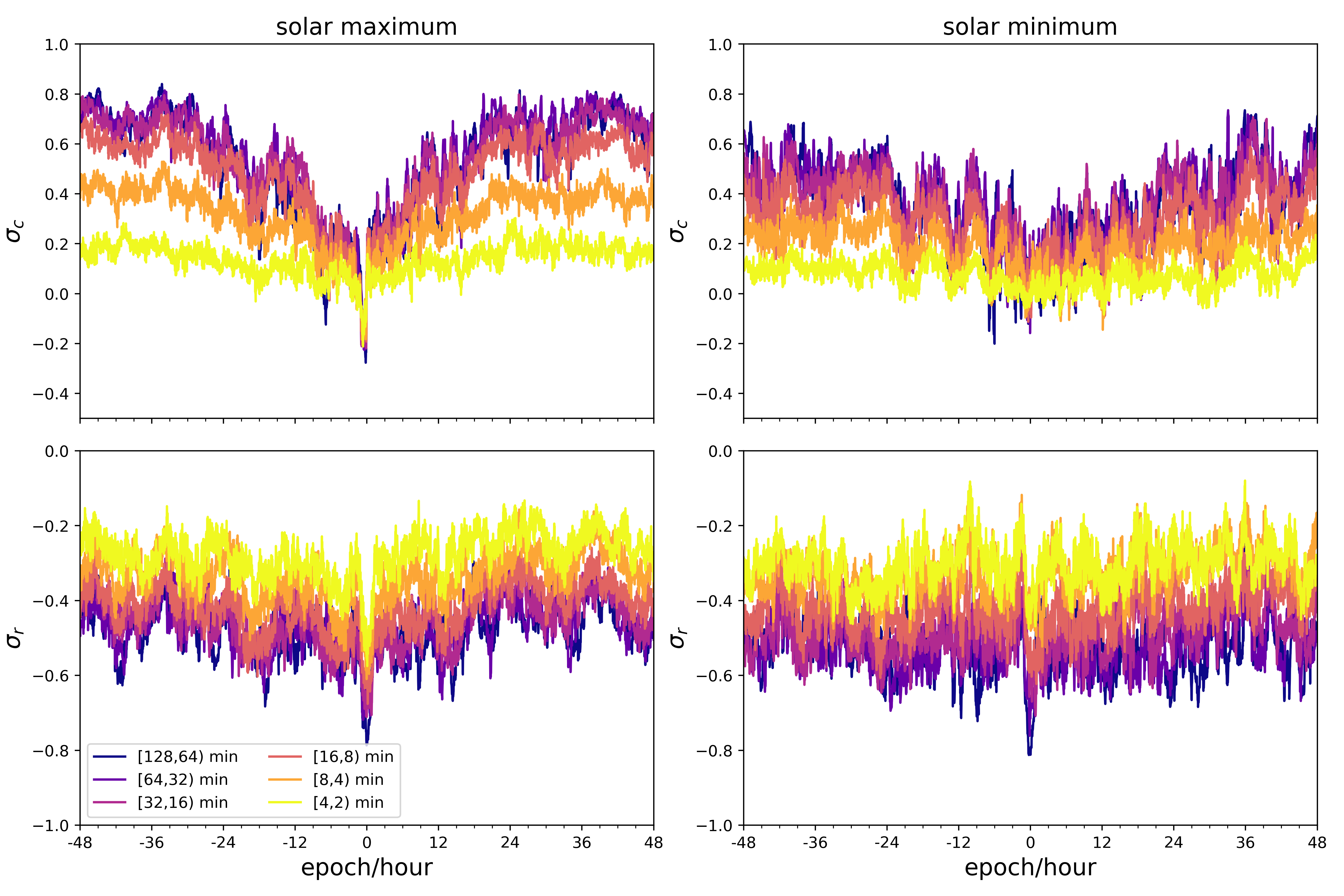}
    \caption{Superposed-epoch analysis of $\sigma_c$ (top row) and $\sigma_r$ (bottom row) near HCS. Left column is solar maximum and right column is solar minimum. In each panel, dark to light colors are wave bands 1-6 respectively, i.e. from low to high frequencies.} 
    \label{fig:sigma_c_sigma_r_superposed}
\end{figure*}

We then analyze turbulence properties near the HCS. We use a running time window of width 128-min. A time window with data gap ratio larger than 20\% is not considered. Inside each time window, we first apply linear interpolation to velocity, magnetic field, and proton density to fill the data gaps. Then we calculate Els\"asser variables after determining the polarity of radial magnetic field by averaging $B_x$ in the time window. Finally, we apply Fourier transform to these fields. Similar to the process of simulation data in the prior section, we divide the frequency into 6 bands such that band $i$ contains wave modes $[2^{i-1}, 2^i)$, e.g. wave band 6 contains waves whose periods are between 128/$2^5$=4 min and $128/2^6$=2 min. We calculate $\sigma_c$ and $\sigma_r$ for each wave band by integrating wave energies in each band. In addition, we fit the power spectra and get the spectral slopes for velocity, magnetic field, outward and inward Els\"asser variables.

In Figure \ref{fig:sigma_c_sigma_r_superposed}, we show superposed-epoch analysis of $\sigma_c$ (top row) and $\sigma_r$ (bottom row). Left column is solar maximum and right column is solar minimum. Colors represent different wave bands such that dark to light colors are bands 1-6. We see that in general $\sigma_c$ decreases with frequency while $\sigma_r$ increases with frequency. Top-left panel shows that in a time window $\pm$ 1 day, approximately the width of the plasma sheet, $\sigma_c$ drops as we approach the center of HCS. In a narrow window of width comparable to the thickness of HCS, i.e. 1-2 hours, $\sigma_c$ drops significantly since the outward and inward waves mix with each other. Bottom-left panel of Figure \ref{fig:sigma_c_sigma_r_superposed} shows slightly decrease of $\sigma_r$ inside the plasma sheet while a large drop of $\sigma_r$ is observed near HCS. In solar minimum (right column of Figure \ref{fig:sigma_c_sigma_r_superposed}), the above results qualitatively hold but both $\sigma_c$ and $\sigma_r$ are lower compared with solar maximum. 

\begin{figure*}[ht!]
    \centering
    \includegraphics[scale=0.5]{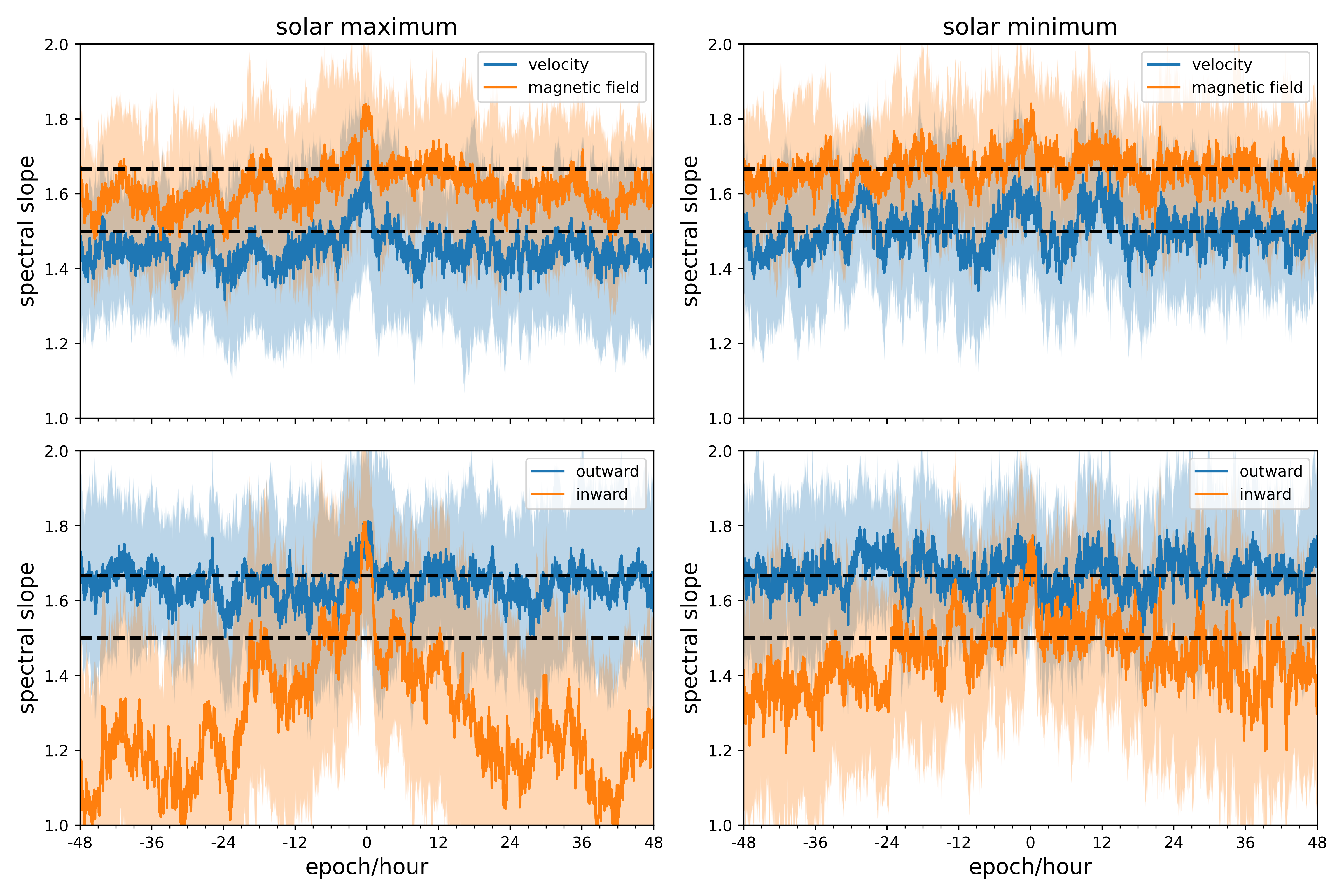}
    \caption{Superposed-epoch analysis of various spectral slopes near HCS. Top panels show slopes of velocity (blue) and magnetic field (orange) and bottom panels show slopes of outward (blue) and inward (orange) Els\"asser variables. Left column is solar maximum and right column is solar minimum. In each panel, the two horizontal lines mark the values $3/2$ and $5/3$ for reference. The shaded bands show the standard deviations of different curves.} 
    \label{fig:spectral_slopes_superposed}
\end{figure*}

In Figure \ref{fig:spectral_slopes_superposed}, we show the superposed-epoch analysis of various spectral slopes. Again, left and right columns are solar maximum and solar minimum respectively. In the top row, blue and orange curves are spectral slopes of velocity and magnetic field. In the bottom row, blue and orange curves are spectral slopes of outward and inward Els\"asser variables. In each panel, the two horizontal dashed lines mark the values $5/3$ and $3/2$ for reference. Away from the HCS, velocity spectrum has a slope near $-3/2$ and magnetic field spectrum has a slope near $-5/3$ as typically observed in the solar wind \citep[e.g.][]{Chenetal2020,shi2021alfvenic}. Near the center of HCS, both of the two spectra steepen and the steepening is more pronounced in solar maximum. The outward Els\"asser variable has a spectral slope near $-5/3$ and shows a slight steepening near HCS in solar maximum while no obvious steepening is observed in solar minimum. The variation in spectral slope of inward Els\"asser variable is more dramatic compared with the other quantities. $\mathbf{z_{in}}$ spectrum is quite flat far from the HCS, around $-1$ in solar maximum and $-1.2$ in solar minimum. We note that, in our simulation (top-left panel of Figure \ref{fig:spectra_sim}), a flatter $\mathbf{z_{in}}$ spectrum compared with the $\mathbf{z_{out}}$ spectrum is also observed.
At the center of HCS, the inward and outward Els\"asser variables have the same slope as expected since the two wave populations are not well separated near the polarity-reversal time.

\begin{figure}
    \centering
    \includegraphics[width=\hsize]{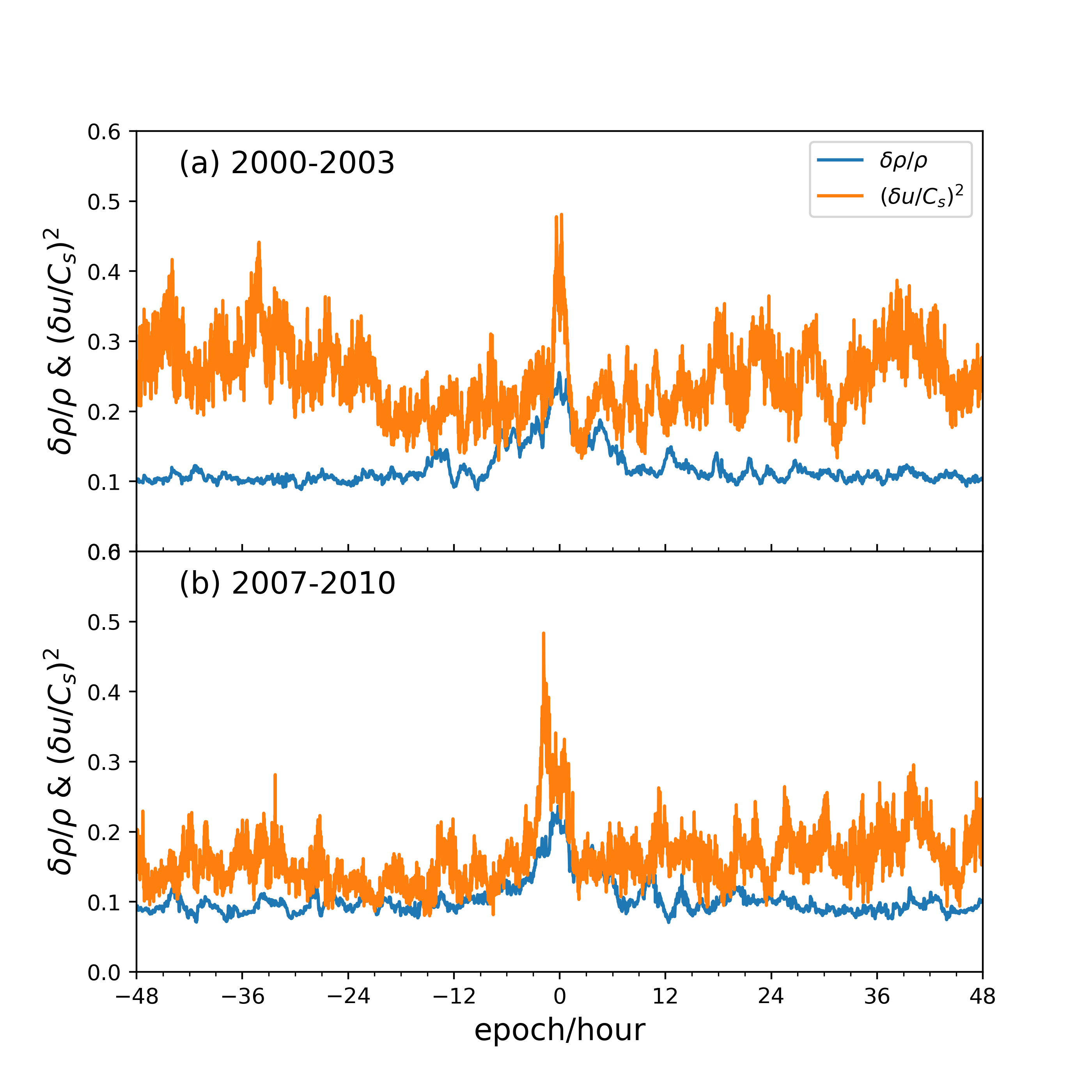}
    \caption{Superposed-epoch analysis of the normalized density fluctuation $\delta \rho / \rho$ (blue) compared with the square of the velocity fluctuation Mach number $(\delta u / C_s)^2$ (orange). Top panel is solar maximum and bottom panel is solar minimum.}
    \label{fig:density_fluctuation_sup_ep}
\end{figure}

In Figure \ref{fig:density_fluctuation_sup_ep}, we plot the superposed-epoch analysis of the normalized density fluctuation $\delta \rho / \rho$ (blue) and the square of the velocity fluctuation Mach number $(\delta u / C_s)^2$ (orange). In solar minimum, the two quantities are highly correlated while in solar maximum they are correlated only close to the HCS. We note that our simulation result (Figure \ref{fig:delta_rho_mach_number_sim}) also shows high correlation between the two parameters close to the current sheets. However, in the simulation, the density fluctuation is smaller near the HCS compared with the surroundings, in contrast to the observation. The reason is that in the initial condition of the simulation we artificially decrease the perturbation amplitude close to the HCS to avoid discontinuity in the perturbation fields. The increase of the density fluctuation near the HCS from OMNI data might be related to the tearing instability of HCS close to the Sun which ejects bunches of flux ropes from the tip of helmet streamers \citep{reville2020tearing}. 

\section{Discussion}\label{sec:discussion}
We compare the simulation results from Section \ref{sec:EBM_sim} and the superposed-epoch analysis from Section \ref{sec:sup_ep_ana}. In the simulation, $\sigma_c$ drops in a wide longitudinal range around the current sheet (Figure \ref{fig:sigma_c_sigma_r_sim}), which is also observed at 1 AU (Figure \ref{fig:sigma_c_sigma_r_superposed}). 
Similarly, in both simulation and observation, $\sigma_r$ drops in the neighbourhood of HCS. \citet{Grappinetal1991} analyzed four-month of Helios 1 data and found that within the neutral sheet, the turbulence properties are close to the ``standard'', or fully-developed, MHD turbulence, rather than Alfv\'enic turbulence. Standard MHD turbulence is characterized by balanced outward/inward Els\"asser energies and an excess of magnetic energy, consistent with our results. In this scenario, the background magnetic field dissipates the residual energy (the so-called ``Alfv\'en effect'') which is a correlation between the two Els\"asser variables generated by intrinsic nonlinear interaction. In other words, the Alfv\'en effect is essentially the dissipation of two colliding (correlated) counter-propagating Alfv\'en wave packets as first described by \citet{kraichnan1965inertial}, and hence it is determined by the background magnetic field strength along the wave propagation direction. Thus, the absolute value of the residual energy, regardless of its sign, is larger inside current sheets where the Alfv\'en effect is weaker. \citet{Grappinetal1991} explained the balance between outward/inward Els\"asser energies at small scales by the fact that the injected energy at large scales due to velocity shear is balanced in $\mathbf{z_{in}}$ and $\mathbf{z_{out}}$. This interpretation, however, cannot explain the decrease in $\sigma_c$ around the current sheet in our simulation because there is no such energy source near the current sheet in the simulation. Instead, the decrease of $\sigma_c$ around the HCS in the simulation is likely to be a result of the shear of background magnetic field which deforms the wave fronts and facilitates the dissipation of wave energies, similar to the velocity shear effect. In addition, \citet{Grappinetal1991} does not answer the question why the residual energy is negative instead of positive in the current sheets. Here we propose a mechanism related to the expansion effect of the solar wind. Near the HCS, the weak background magnetic field allows fluctuations to evolve freely so they are dominated by the spherical expansion effect, which leads to $\mathbf{u_\perp}, \mathbf{b_\perp} \sim 1/R$, $ \rho \sim 1/R^2$ and consequently $\mathbf{b_\perp}/\sqrt{\rho} \sim 1$.  Hence, as the radial distance increases, the transverse magnetic field fluctuation (in Alfv\'en speed) becomes larger than the transverse velocity fluctuation, leading to a negative $\sigma_r$. This mechanism is supported by Figure \ref{fig:sigma_c_sigma_r_radius_CS} which shows that without expansion no net residual energy is produced. Meanwhile, Figure \ref{fig:sigma_c_sigma_r_radius_CS} also shows that the decrease of $\sigma_c$ cannot be explained by expansion effect and must be caused by processes related to the shear of the background magnetic field. 

We note that, our simulation cannot explain why in the solar wind $\sigma_r$ is generally negative even far from HCS as can be seen from Figure \ref{fig:sigma_c_sigma_r_superposed}. Recent studies using Parker Solar Probe data show that $\sigma_r$ is already negative at below 30 solar radii while $\sigma_c$ is increasingly high as the satellite moves closer to the Sun \citep{Chenetal2020,shi2021alfvenic}. Our results show that the presence of a current sheet indeed leads to a dominance of magnetic energy, but it also results in a decrease in $\sigma_c$. Thus the observed $(\sigma_c \approx 0, \sigma_r \approx -1)$ population of the solar wind fluctuations \citep[e.g.][]{DAmicisandBruno2015} is possibly Alfv\'enic turbulence evolved under the influence of current sheets, while the prevailing $(\sigma_c \lesssim 1, \sigma_r \approx -0.2)$ population may be generated in the very young solar wind with other processes taking effect or it may be a natural result of the evolutin of Alfv\'enic turbulence \citep[e.g.][]{boldyrev2011residual}.

Last, we would like to comment that, the statistical study of Alfv\'enic turbulence properties near SIRs by \citet{BorovskyandDenton2010} shows results quite different from our simulations. Their Figure 11 and Figure 16 show that, at the fast-slow stream interface, the magnetic energy dominance is enhanced, i.e. $\sigma_r$ decreases, and the Els\"asser ratio $|\mathbf{z_{out}}|/|\mathbf{z_{in}}|$ increases, contradicting our simulation results that $\sigma_c$ declines and $\sigma_r$ increases at SIRs (Figure \ref{fig:sigma_c_sigma_r_sim}). The reason for this contradiction is unknown and needs further study.

\section{Conclusion}\label{sec:conclusion}
In this study, we carry out two-dimensional MHD simulations, using expanding-box-model, and a superposed-epoch analysis, using OMNI data, to study the turbulence properties in the solar wind with a focus on the heliospheric current sheet. The simulation results show that both the normalized cross helicity $\sigma_c$ and normalized residual energy $\sigma_r$ drop in the neighborhood of HCS (Figure \ref{fig:sigma_c_sigma_r_sim}). The observation at 1 AU shows that $\sigma_c$ and $\sigma_r$ decrease sharply at the center of HCS, on a time scale of $1-2$ hours which is the scale of the HCS crossings (Figure \ref{fig:sigma_c_sigma_r_superposed}). The observation also shows that $\sigma_c$ starts to drop gradually in a much wider time range $\Delta t > \pm 1$ day, inside the plasma sheet bounding the HCS. The power spectra, calculated over frequency range $f \in [128^{-1}, 2^{-1}]$ min$^{-1}$ using OMNI data, of velocity, magnetic field, outward and inward Els\"asser variables steepen near the HCS (Figure \ref{fig:spectral_slopes_superposed}), and steeper parallel power spectra near the HCS are also observed in the simulations (Figure \ref{fig:spectra_sim}), implying a stronger energy cascade of the turbulence. 
Last, both the simulation (Figure \ref{fig:delta_rho_mach_number_sim}) and the satellite observation (Figure \ref{fig:density_fluctuation_sup_ep}) show that around the HCS, the density fluctuation $\delta \rho/\rho$ is highly correlated with the square of the velocity fluctuation Mach number $(\delta u/C_s)^2$, implying a significant compressive component in the velocity fluctuations near the HCS \citep{Grappinetal1991}.

Our results confirm that current sheets significantly influence the evolution of solar wind turbulence in a different way from the velocity shear as discussed by \citet{Shietal2020}. They may explain the low cross helicity and magnetic-energy-dominated population of fluctuations in the solar wind. But the origin of the prevailing high cross helicity and slightly magnetic-energy-dominated fluctuations needs other mechanisms that play important roles close to the Sun, or at the source region of the Alfv\'enic fluctuations. Inspection of the Parker Solar Probe data is necessary to fully understand these mechanisms.

\acknowledgments
The OMNI data were obtained from the GSFC/SPDF OMNIWeb interface at https://omniweb.gsfc.nasa.gov. This work used the Extreme Science and Engineering Discovery Environment (XSEDE) EXPANSE at SDSC through allocation No. TG-AST200031, which is supported by National Science Foundation grant number ACI-1548562 \citep{Townsetal2014}. The work was supported by NASA HERMES DRIVE Science Center grant No. 80NSSC20K0604.

\software{Matplotlib \citep{Hunter2007Matplotlib}}

%

\appendix
\section{List of heliospheric current sheet crossings identified using OMNI data}
The full list of the HCS-crossings is shown in Table \ref{tab:HCS_crossing}.
\begin{table}[ht!]
    \caption{List of HCS-crossings identified using OMNI data}
 \begin{minipage}{0.5\linewidth}
     \begin{tabular}{c | c c c c c} 
     \hline
     \# & year & month & day & hour & minute \\
     \hline
     01 & 2000 & 01 & 10 & 00 & 30 \\
     02 & 2000 & 02 & 05 & 17 & 50 \\
     03 & 2000 & 07 & 31 & 19 & 40 \\
     04 & 2000 & 08 & 27 & 17 & 33 \\
     05 & 2000 & 09 & 24 & 15 & 50 \\
     06 & 2000 & 10 & 14 & 18 & 04 \\
     07 & 2000 & 11 & 23 & 19 & 33 \\
     08 & 2000 & 12 & 16 & 21 & 03 \\
     09 & 2000 & 12 & 22 & 21 & 23 \\
     10 & 2001 & 01 & 10 & 21 & 03 \\
     11 & 2001 & 02 & 14 & 07 & 17 \\
     12 & 2001 & 03 & 12 & 14 & 55 \\
     13 & 2001 & 04 & 22 & 00 & 23 \\
     14 & 2001 & 05 & 06 & 10 & 40 \\
     15 & 2001 & 05 & 17 & 21 & 32 \\
     16 & 2001 & 06 & 29 & 06 & 21 \\
     17 & 2001 & 07 & 10 & 16 & 30 \\
     18 & 2001 & 07 & 24 & 15 & 05 \\
     19 & 2001 & 11 & 16 & 11 & 28 \\
     20 & 2002 & 02 & 04 & 21 & 21 \\
     21 & 2002 & 03 & 03 & 22 & 49 \\
     22 & 2002 & 05 & 06 & 09 & 55 \\
     23 & 2002 & 06 & 02 & 02 & 40 \\
     24 & 2002 & 06 & 16 & 06 & 08 \\
     25 & 2002 & 06 & 25 & 16 & 37 \\
     26 & 2002 & 09 & 03 & 06 & 46 \\
     27 & 2002 & 09 & 27 & 05 & 29 \\ 
     28 & 2002 & 10 & 23 & 17 & 02 \\
     29 & 2002 & 11 & 10 & 02 & 57 \\ 
     30 & 2002 & 12 & 06 & 11 & 21 \\ 
     31 & 2002 & 12 & 19 & 07 & 41 \\
     32 & 2003 & 01 & 17 & 14 & 08 \\
     33 & 2003 & 02 & 12 & 22 & 54 \\
     34 & 2003 & 02 & 26 & 19 & 48 \\
     35 & 2003 & 03 & 11 & 17 & 18 \\
     36 & 2003 & 03 & 26 & 09 & 40 \\
     37 & 2003 & 04 & 08 & 02 & 34 \\
     38 & 2003 & 04 & 20 & 19 & 07 \\
     39 & 2003 & 05 & 04 & 16 & 00 \\
     40 & 2003 & 05 & 18 & 16 & 23 \\
     41 & 2003 & 06 & 26 & 12 & 30 \\
     42 & 2003 & 07 & 11 & 15 & 25 \\
     43 & 2003 & 07 & 26 & 12 & 01 \\
     44 & 2003 & 08 & 04 & 06 & 52 \\
     45 & 2003 & 09 & 01 & 06 & 13 \\
     46 & 2003 & 10 & 13 & 09 & 27 \\
     47 & 2003 & 12 & 05 & 01 & 26 \\ 
     48 & 2003 & 12 & 19 & 19 & 50 \\
     \hline
     \end{tabular}
\end{minipage} 
 \begin{minipage}{0.5\linewidth}
     \begin{tabular}{c | c c c c c} 
     \hline
     \# & year & month & day & hour & minute \\
     \hline
     01 & 2007 & 01 & 08 & 02 & 01 \\
     02 & 2007 & 01 & 15 & 08 & 39 \\
     03 & 2007 & 02 & 04 & 01 & 44 \\
     04 & 2007 & 02 & 12 & 15 & 17 \\
     05 & 2007 & 03 & 03 & 08 & 18 \\
     06 & 2007 & 03 & 11 & 18 & 17 \\
     07 & 2007 & 03 & 31 & 23 & 21 \\
     08 & 2007 & 06 & 02 & 15 & 19 \\
     09 & 2007 & 06 & 08 & 01 & 24 \\
     10 & 2007 & 06 & 13 & 18 & 59 \\
     11 & 2007 & 08 & 05 & 16 & 02 \\
     12 & 2007 & 08 & 31 & 20 & 25 \\
     13 & 2007 & 09 & 09 & 23 & 45 \\
     14 & 2007 & 10 & 11 & 06 & 22 \\
     15 & 2007 & 11 & 20 & 09 & 33 \\
     16 & 2007 & 12 & 17 & 06 & 12 \\ 
     17 & 2008 & 01 & 12 & 13 & 06 \\
     18 & 2008 & 01 & 31 & 15 & 24 \\
     19 & 2008 & 02 & 07 & 17 & 25 \\
     20 & 2008 & 02 & 27 & 17 & 03 \\
     21 & 2008 & 03 & 08 & 08 & 01 \\
     22 & 2008 & 04 & 03 & 03 & 00 \\
     23 & 2008 & 04 & 22 & 15 & 28 \\
     24 & 2008 & 04 & 30 & 16 & 56 \\
     25 & 2008 & 06 & 25 & 16 & 14 \\
     26 & 2008 & 07 & 21 & 03 & 41 \\
     27 & 2008 & 07 & 30 & 10 & 50 \\
     28 & 2008 & 08 & 24 & 23 & 43 \\
     29 & 2008 & 11 & 07 & 03 & 51 \\
     30 & 2008 & 12 & 03 & 15 & 40 \\
     31 & 2008 & 12 & 11 & 00 & 09 \\
     32 & 2009 & 01 & 23 & 13 & 54 \\
     33 & 2009 & 02 & 03 & 00 & 59 \\
     34 & 2009 & 04 & 15 & 08 & 27 \\
     35 & 2009 & 05 & 13 & 19 & 36 \\
     36 & 2009 & 06 & 20 & 19 & 18 \\
     37 & 2009 & 07 & 13 & 13 & 23 \\
     38 & 2009 & 07 & 20 & 10 & 21 \\
     39 & 2010 & 01 & 31 & 02 & 21 \\
     40 & 2010 & 03 & 01 & 06 & 52 \\
     41 & 2010 & 03 & 14 & 23 & 10 \\
     42 & 2010 & 06 & 06 & 23 & 58 \\
     43 & 2010 & 08 & 20 & 10 & 01 \\
     44 & 2010 & 11 & 12 & 11 & 26 \\
     45 & 2010 & 12 & 23 & 16 & 26 \\
     & & & & & \\
     & & & & & \\
     & & & & & \\
     \hline
     \end{tabular}
 \end{minipage}\label{tab:HCS_crossing}
\end{table}





\bibliography{bibChenShi}{}
\bibliographystyle{aasjournal}




\end{CJK*}
\end{document}